\documentclass[aps,pre,10pt,twocolumn,showpacs,final,letterpaper]{revtex4-1}
\usepackage{amsmath,amssymb}
\usepackage{graphicx}
\usepackage{subfigure}
\usepackage[pdftex]{hyperref}
\usepackage{bm}
\usepackage{algcompatible}

\begin{document}
\title{Percolation on random networks with arbitrary k-core structure}
\author{Laurent \surname{H\'ebert-Dufresne}}\thanks{These two authors contributed equally to this work.}
\author{Antoine \surname{Allard}}\thanks{These two authors contributed equally to this work.}
\author{Jean-Gabriel \surname{Young}}
\author{Louis J. \surname{Dub\'e}}
\affiliation{D\'epartement de Physique, de G\'enie Physique, et d'Optique, Universit\'e Laval, Qu\'ebec (Qu{\'e}bec), Canada G1V 0A6}
\date{\today}
\pacs{64.60.aq,64.60.ah}
%
%
%
%
%
\begin{abstract}
 The k-core decomposition of a network has thus far mainly served as a powerful tool for the empirical study of complex networks. We now propose its explicit integration in a theoretical model. We introduce a Hard-core Random Network model that generates maximally random networks with arbitrary degree distribution \textit{and} arbitrary k-core structure. We then solve exactly the bond percolation problem on the HRN model and produce fast and precise analytical estimates for the corresponding real networks. Extensive comparison with real databases reveals that our approach performs better than existing models, while requiring less input information.
\end{abstract}
\maketitle 
%
%
%
%
%
\section{Introduction}
%

We address the challenge of designing a realistic model of complex networks while preserving its analytic tractability. The model should include the essential structural properties of real networks, and the theoretical framework should guarantee easy access to quantitative calculations. For the second aspect of this endeavour, we cast our analysis in terms of a percolation problem. This has been a topic of choice for some years since it can just as well represent the dynamics \textit{of} a network as the dynamics \textit{on} the network \cite{Dorogovtsev03_Evolution,Meyers07_BullAmerMathSoc,Arenas08_PhysRep,Dorogovtsev08_RevModPhys,Cohen10_ComplexNetworks,Newman10_Networks,Hebert-Dufresne13_PhysRevLett,Hebert-Dufresne13_SciRep}. One might think of its growth, its robustness (to attacks or failures) and the propagation of emerging infectious agents (e.g. disease or information).

While the study of percolation models on idealized networks has led to a better understanding of both the processes they model and the networks that support them, the study of percolation on real networks has somewhat stagnated. Unfortunately, purely numerical approaches are time-consuming, require a complete description of the networks under scrutiny and lack the insights of an analytical description. Conversely, although analytical modeling provides a better understanding of the organization of real networks, they are limited at present to simplified random models \cite[see][{and references therein}]{Newman03_SIAMRev,Newman10_Networks}.

In this paper, we demonstrate how the k-core structure of networks (hereafter simply core structure) plays a central role in the outcome of bond percolation, and how it acts as a proxy that captures the essential structural properties of real networks. The ensuing model, that we call the Hard-core Random Network (HRN) model, creates maximally random networks with an arbitrary degree distribution \textit{and} an arbitrary core structure. We also propose a Metropolis-Hastings algorithm to generate such random networks. The HRN model serves our purpose well since it is shown to be amenable to an exact solution for the size of the extensive ``giant'' component (in the limit of large network size).  With less input information, it outperforms the current standard model \cite{Melnik11_PhysRevE} for precise prediction of percolation results on real networks.

The organization of this paper goes as follows. In Sec.~\ref{sec:hrn_perco}, we introduce the bond percolation problem and briefly present the two models used for comparison. In Sec.~\ref{sec:hrn_hrn}, we present the HRN model, the equations used to solve the bond percolation problem and the Metropolis-Hastings algorithm generating the corresponding random networks. We also compare the predictions of the HRN model and the ones of the two aforementioned models with the results obtained numerically using real network databases. Final remarks are collected in the last section.
%
%
%
%
%
\section{Bond percolation on networks} \label{sec:hrn_perco}
%

The bond percolation problem concerns the connectivity of a network after the removal of a fraction $1-T$ of its edges. More precisely, for a synthetic or empirical network, we are interested in the fraction $S$ of nodes contained in the largest connected component---the giant component---after each edge has been removed independently with a probability $1-T$. In the limit of large networks, this component undergoes a \textit{phase transition} at a critical point $T_\mathrm{c}$ during which its size (the number of nodes it contains) becomes an extensive quantity that scales linearly with the number of nodes ($N$) of the whole network \cite{Christensen05_ComplexityAndCriticality}.

\begin{figure}[t]
  \subfigure[\ CM]{\label{fig:hrn_example_CM}   \includegraphics[width=0.25\linewidth]{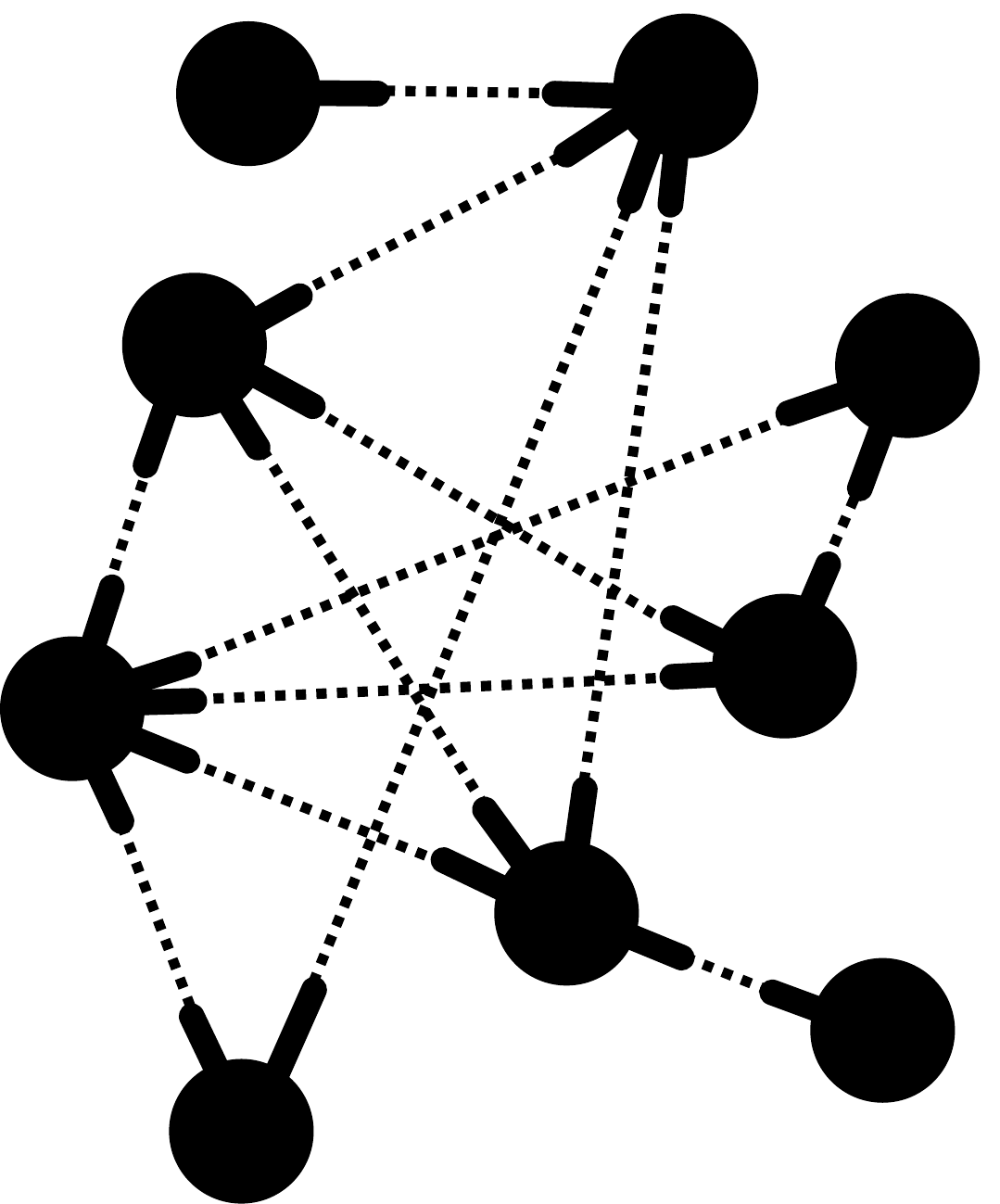}}
  \hspace{0.05\linewidth}
  \subfigure[\ CCM]{\label{fig:hrn_example_CCM} \includegraphics[width=0.25\linewidth]{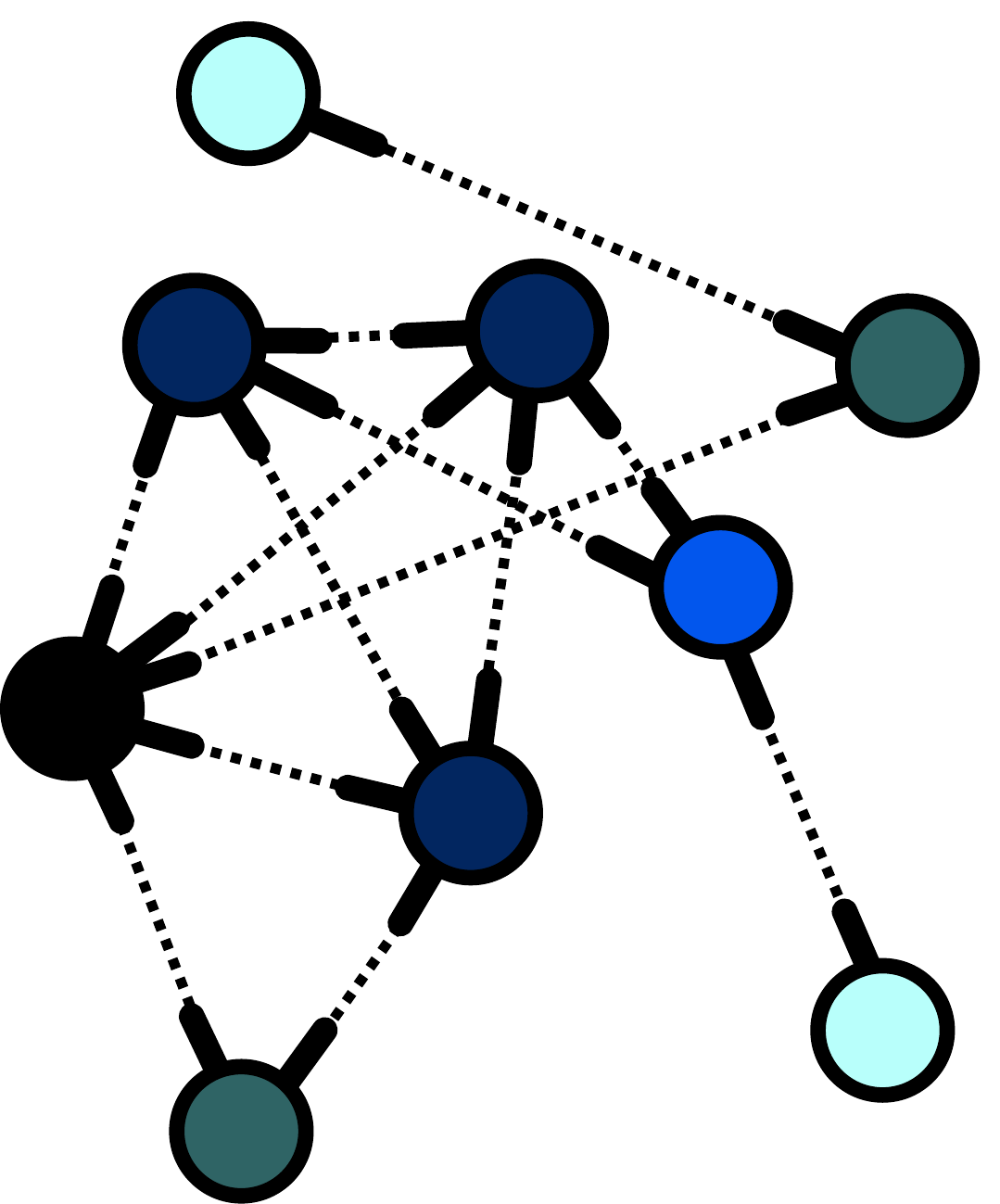}}
  \hspace{0.05\linewidth}
  \subfigure[\ HRN]{\label{fig:hrn_example_HRN} \includegraphics[width=0.25\linewidth]{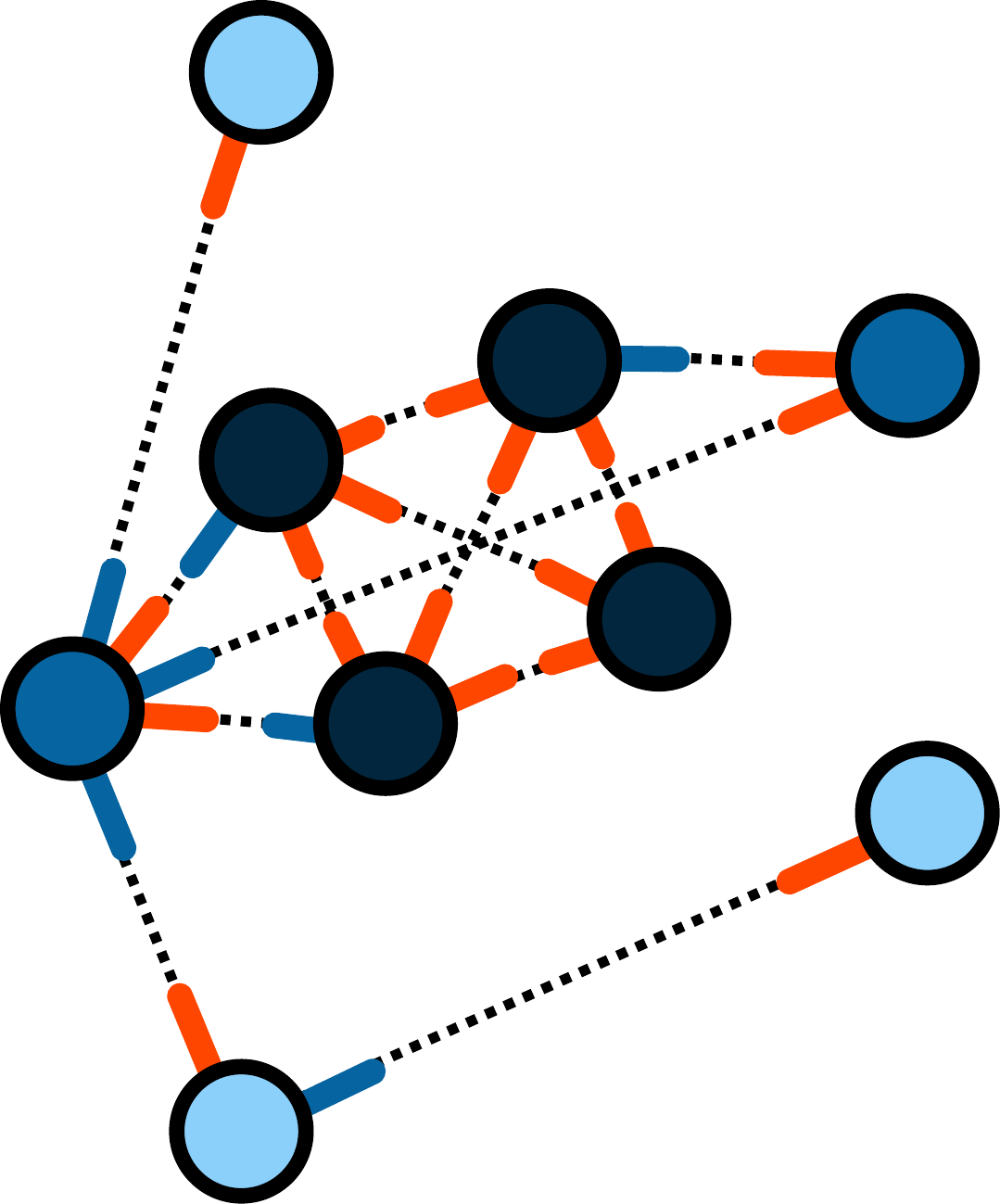}}
  \caption{\label{fig:hrn_example}(color online). Comparison of the three random network models considered. (a) The CM randomly connects stubs drawn from a given degree distribution $\{P(k)\}_{k\in\mathbb{N}}$. (b) The CCM distinguishes nodes according to their degree (colors) and randomly match stubs according to the joint degree distribution $\{P(k,k^\prime)\}_{k,k^\prime\in\mathbb{N}}$. (c) The HRN model distinguishes nodes by their coreness (colors) and stubs by their contribution to a node's coreness (red or blue). Stubs are then randomly matched according to the matrices $\mathbf{K}$ and $\mathbf{C}$.}
\end{figure}

To compare and assert the precision of the predictions of our model, we use the \textit{Configuration Model} (CM) and \textit{Correlated Configuration Model} (CCM) \cite{Newman01_PhysRevE,Newman02_PhysRevE,Newman02_PhysRevLett,Vazquez03_PhysRevE} (see Fig.~\ref{fig:hrn_example_CM}--\subref{fig:hrn_example_CCM}) as benchmarks. These models define maximally random network ensembles that are random in all respects other than the degree distribution (CM,CCM) and the degree-degree correlations (CCM). The degree distribution, $\{P(k)\}_{k\in\mathbb{N}}$, is the distribution of the number of connections (the degree $k$) that nodes have. The degree-degree correlations are defined through the \textit{joint degree distribution}, $\{P(k,k^\prime)\}_{k,k^\prime\in\mathbb{N}}$, giving the probability that a randomly chosen edge has nodes of degree $k$ and $k^\prime$ at its ends.

For both models, the size of the giant component $S$ and the percolation threshold $T_\mathrm{c}$ can be calculated in the limit $N\rightarrow\infty$ using probability generating functions (pgf) \cite{Newman01_PhysRevE,Newman02_PhysRevE,Newman02_PhysRevLett,Vazquez03_PhysRevE,Newman03a_PhysRevE,Vazquez06_PhysRevE,Allard09_PhysRevE,Allard12_JPhysA}. To model bond percolation on a given network with these models, we simply extract the degree distribution and the joint degree distribution; the required information therefore scales as $k_{\mathrm{max}}$ and $k_{\mathrm{max}}^2$. The original network is then found within the random ensembles containing all possible networks that can be designed with the same degree distribution and/or degree-degree correlations. The readers unfamiliar with these models and/or the mathematics involved can get a brief overview of these subjects in Appendices \ref{app:CM} and \ref{app:CCM}.

The degree distribution and the joint degree distribution can be seen as the one-point and two-point correlation functions of a network. The next logical step would therefore be to consider three-point correlations (i.e., clustering), and eventually to incorporate mesoscopic features such as motifs, cliques, and communities. Although many theoretical models have been proposed \cite{Newman03b_PhysRevE,Serrano06_PhysRevLett,Serrano06b_PhysRevE,Shi07_PhysicaA,Berchenko09_PhysRevLett,Miller09_PhysRevE,Newman09_PhysRevLett,Gleeson09_PhysRevE,Karrer10_PhysRevE,Zlatic12_EPL,Allard12_JPhysA}, a general, objective, and systematic method to tune these models in order to reproduce the features found in real networks as well as to predict the outcome of bond percolation is yet to be found \footnote{Recent advances in understanding the global organization of clustering in real networks \cite{Colomer-de-Simon13} offers further ideas to incorporate clustering in our model and will be the subject of a subsequent study.}.
%
%
%
%
%
\section{Hard-core Random Networks (HRN)} \label{sec:hrn_hrn}
%
We propose an alternative approach by considering a macroscopic measure of centrality: the \textit{coreness} of nodes. This choice is motivated by the recent observation that a node's coreness is a better indicator of the likeliness for that node to be part of the giant component than its degree \cite{Kitsak10_NaturePhys}. This measure also has the advantage of being general, objective, systematic, and easily calculated \cite{Batagelj03_arXiv}.

\subsection{Network coreness}

The coreness $c$ of a node is specified through its position in the core decomposition of a network \cite{Seidman83_SocialNetworks,Dorogovtsev06_PhysRevLett}. This decomposition assigns nodes to nested cores where nodes belonging to the $n$-th core all share at least $n$ edges with one another. A node has a coreness equal to $c$ if it is found in the $c$-th core, but not in the $(c+1)$-th core. The set of nodes with a coreness equal to $c$ forms the $c$-shell.

This definition of the coreness may appear complicated to compute, but a simple algorithm allows us to do the decomposition very efficiently \cite{Batagelj03_arXiv}.
\algblock{If}{EndIf}
\algcblock[If]{If}{ElsIf}{EndIf}
\algcblock{If}{Else}{EndIf}
\algcblockdefx[Strange]{}{Eeee}{Oooo}
[1]{\textbf{Input} #1}
[1]{\textbf{Output} #1}
\begin{algorithmic}[1]
\Eeee{graph as lists of nodes $\mathcal{V}$ and neighbors $\mathcal{N}$}
\Oooo{list $\mathcal{C}$ with coreness for each node}
\STATE{compute and list the degrees $\mathcal{D}$ of nodes;}
\STATE{sort $\mathcal{V}$ with increasing degree of nodes;}
\FORALL{$v \in \mathcal{V}$ in the order of $\mathcal{V}$}
	\STATE{$\mathcal{C}(v)$ := $\mathcal{D}(v)$;} 
    \FORALL{$u \in \mathcal{N}(v)$}
    	\IF{$\mathcal{D}(u) > \mathcal{D}(v)$}
        	\STATE{$\mathcal{D}(u)$ := $\mathcal{D}(u) -1$;}
        \ENDIF
    \ENDFOR
  \STATE{re-sort $\mathcal{V}$ accordingly}
\ENDFOR
\end{algorithmic}
In short, this algorithm is similar to a \textit{pruning} process which removes nodes in order of their effective degree, i.e., their number of links shared with nodes currently ranked higher in the process. In the end, the coreness of a node is simply given by its degree once the peeling process reaches this particular node. Hence, we know that a node of degree $k$ and coreness $c$ has $c$ \textit{contributing} edges and $k-c$ \textit{non-contributing} edges. Based on this key observation, we develop a coreness-based random network model that defines a maximally random network ensemble with an arbitrary degree distribution \textit{and} an arbitrary core structure.
\subsection{The HRN model}
The only two inputs of the HRN model are a $\mathbf{K}$ matrix whose elements $K_{ck}$ correspond to the fraction of the nodes that have a coreness $c$ and a degree $k$, and a matrix $\mathbf{C}$ whose elements $C_{cc^\prime}$ give the fraction of edges that leave nodes of coreness $c$ to nodes of coreness $c^\prime$. As this model considers undirected networks, the matrix $\mathbf{C}$ is symmetric and each edge is counted twice to account for both directions.

The HRN model is a multitype version of the CM \cite{Allard09_PhysRevE,Allard12_JPhysA,Allard13b} in which each node is assigned to a type, its coreness, and in which edges are formed by randomly pairing stubs that either contribute to the node's coreness (say, \textit{red} stubs) or do not contribute to it (say, \textit{blue} stubs). Red stubs from nodes of coreness $c$ may be paired with blue stubs from nodes of coreness $c^\prime \geq c$, or with red stubs attached to nodes of coreness $c^\prime=c$ (intra-shell). Blue stubs stemming from nodes of coreness $c$ may only be matched with red stubs stemming from nodes with a coreness $c^\prime \leq c$. Blue stubs may never be paired together.

These rules enforce a minimal core structure, although random variations can bring nodes to a higher coreness than originally intended. For example, 3 nodes of original state $(k=2,c=1)$ could end up in the 2-shell in the unlikely event that they form a triangle. However, such random variations may never pull nodes to a lower coreness than intended, in addition to being extremely unlikely in the limit of large networks ($N \rightarrow \infty$). The matrices $\mathbf{K}$ and $\mathbf{C}$ (see Appendix \ref{app:cons} for consistency conditions) combined with the aforementioned stub pairing rules define a maximally random network ensemble with an arbitrary degree distribution and core structure (see Fig.~\ref{fig:hrn_example_HRN}).

The $\mathbf{K}$ matrix encodes several useful quantities. For instance, the fraction of nodes of coreness $c$
\begin{align} \label{eq:hrn_wc}
  w_c & = \sum_{k} K_{ck} \ ,
\end{align}
and the associated joint degree distribution, i.e. the probability that a randomly chosen node of coreness $c$ has $k_r$ red stubs and $k_b$ blue stubs
\begin{align}
 P_c(\bm{k}) \equiv P_c(k_r,k_b) = \frac{\delta_{c,k_r}}{w_c}K_{k_r,k_r+k_b} \ ,
\end{align}
where $\delta_{c,k_r}$ is the Kronecker delta. Furthermore, we can extract the average degree of nodes of coreness $c$
\begin{align}
 \langle k \rangle_c = \frac{1}{w_c}\sum_{k}k K_{c,k}
\end{align}
\begin{figure}[t]
  \includegraphics[trim = 0mm 0mm 0mm 0mm, clip, height=4.95cm]{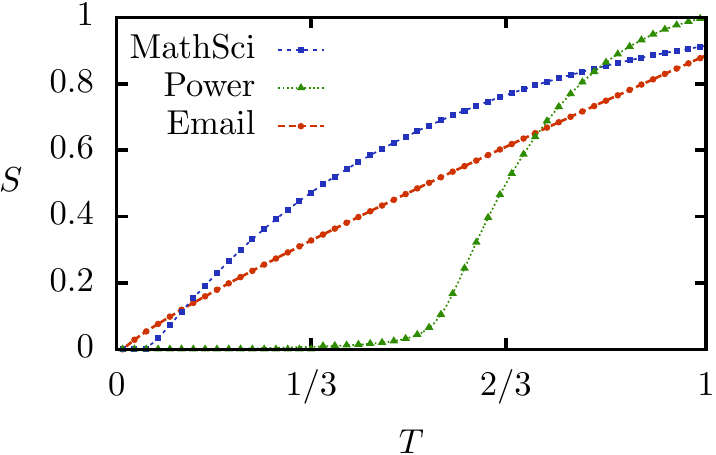}
  \caption{\label{fig:hrn_validation}(color online). Validation of the HRN model. The predictions of Eqs.\eqref{eq:hrn_pgf_order_param}--\eqref{eq:hrn_pgf_fixed_point} (lines) are compared with the results obtained on networks generated with the Metropolis-Hastings algorithm described in Sec.~\ref{sec:hrn_hrn_algo} (symbols). The matrices $\mathbf{K}$ and $\mathbf{C}$ were extracted from an email network, the MathSciNet co-authorship network and a power grid chosen for their different behaviors (see Table~\ref{tab:hrn_networks} for datasets details). Numerical results (symbols) represent the average value of over $5\cdot 10^5$ simulations performed on networks with more than $3\cdot 10^ 5$ nodes.}
\end{figure}
and the average degree of the whole network
\begin{align}
 \langle k \rangle = \sum_{c,k}k K_{ck} \ .
\end{align}
It follows from the above definition that a fraction $w_c\langle k \rangle_c/\langle k \rangle$ of stubs stems from nodes of coreness $c$, of which a fraction $w_cc/\langle k \rangle$ is red and a fraction $w_c(\langle k \rangle_c-c)/\langle k \rangle$ is blue.

The $\mathbf{C}$ matrix encodes the transition probability $R(c^\prime,j|c,i)$ that a node of coreness $c$ through a stub of color $i$ [red ($r$) or blue ($b$)] leads to a node of coreness $c^\prime$ through one of its stubs of color $j$. Since inter-shell edges can only be formed by matching a red with a blue stub, we readily obtain
\begin{subequations} \label{eq:hrn_R}
\begin{align}
  R(c^\prime,b|c,r)   & = \frac{C_{cc^\prime}}{w_cc/\langle k \rangle} \\
  R(c,r|c^{\prime},b) & = \frac{C_{cc^\prime}}{w_c(\langle k \rangle_c - c )/\langle k \rangle} \\
  R(c^\prime,r|c,b)   & = R(c,b|c^{\prime},r) = 0
\end{align}
for $c < c^\prime$. Similarly, as the pairing of blue stubs is forbidden [$R(c^\prime,b|c,b)=0$ for any $c$ and $c^\prime$], a blue stub stemming from a node of coreness $c$ leads to a node belonging to the same shell (through its red stub) with probability 
\begin{align} \label{eq:hrn_Rcbrc}
  R(c,r|c,b) & = \frac{w_c(\langle k \rangle_c - c )/\langle k \rangle - \sum_{c^{\prime\prime}<c} C_{cc^{\prime\prime}}}{w_c(\langle k \rangle_c - c )/\langle k \rangle} \ .
\end{align}
\begin{figure*}[t!h]
  \includegraphics[trim = 0mm 4mm 0mm 0mm, clip, height=3.95cm]{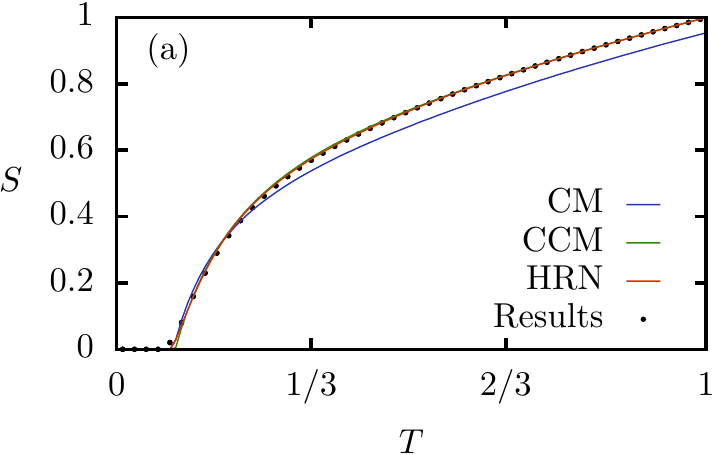}
  \includegraphics[trim = 4mm 4mm 0mm 0mm, clip, height=3.95cm]{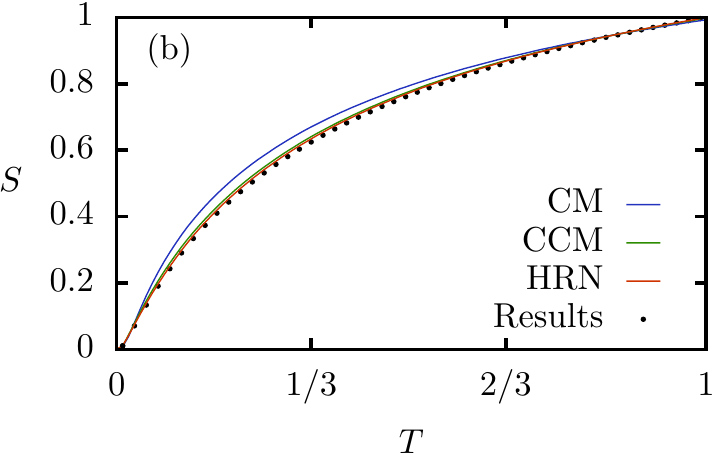} \\
  \includegraphics[trim = 0mm 4mm 0mm 0mm, clip, height=3.95cm]{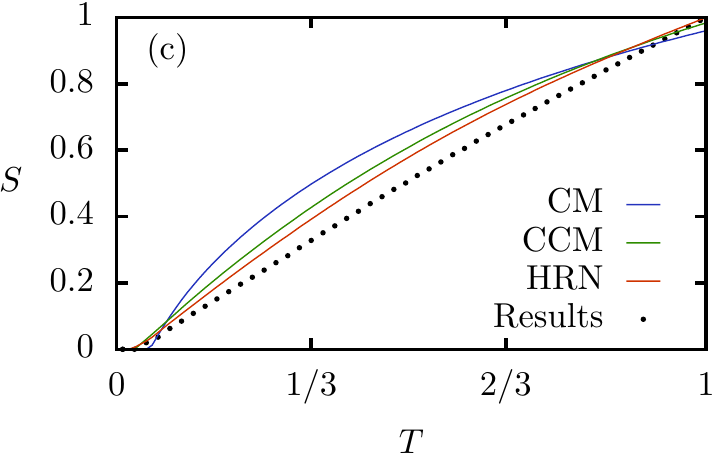}
  \includegraphics[trim = 4mm 4mm 0mm 0mm, clip, height=3.95cm]{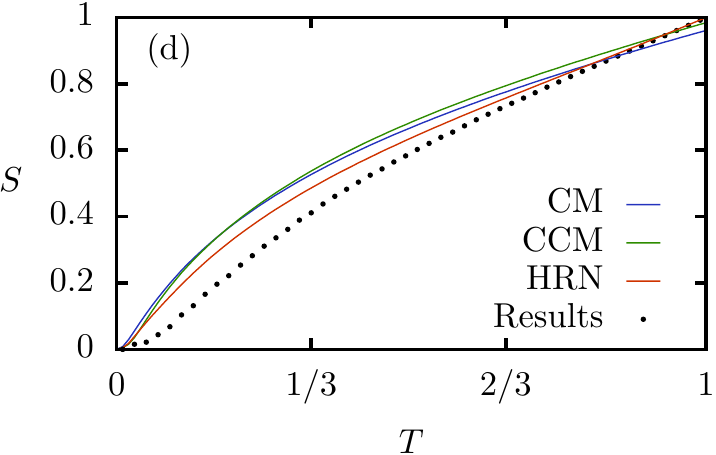}
  \includegraphics[trim = 0mm 0mm 0mm 0mm, clip, height=4.345cm]{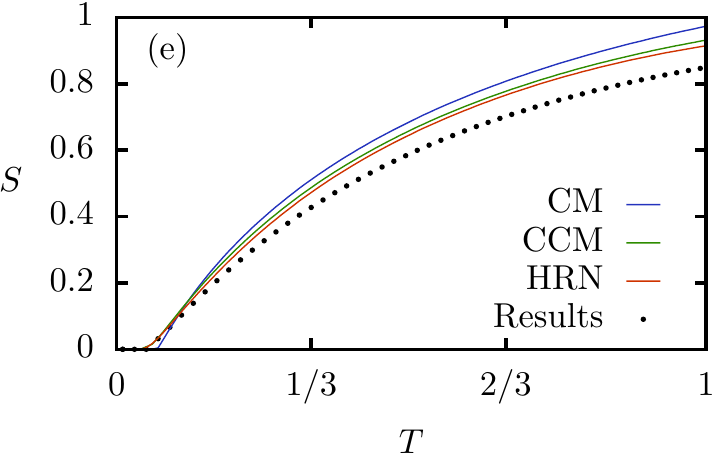}  
  \includegraphics[trim = 4mm 0mm 0mm 0mm, clip, height=4.345cm]{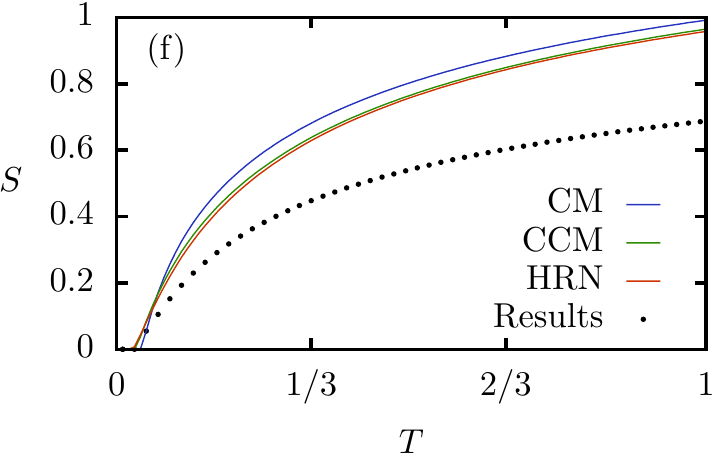} \\
  \caption{\label{fig:hrn_results_1}(color online). Results of bond percolation on real networks (black) compared with analytical predictions obtained with the CM (blue), CCM (green) and HRN (red). The networks are: (a) a snapshot of the Gnutella peer-to-peer network, (b) a snapshot of the Gowalla location-based social network, (c) the Pretty-Good-Privacy trust network, (d) a subset of the World-Wide Web, (e) the co-authorship network of MathSciNet before 2008, and (f) a large subset of the Facebook social network. See Table~\ref{tab:hrn_networks} for further details.}
\end{figure*}
This last result is computed by subtracting the number of blue stubs leading to outer shells (i.e., lower coreness) to the total number of blue stubs stemming from nodes of coreness $c$, and then by normalizing [$\sum_{c^\prime,j}R(c^{\prime},j|c,i)=1$ for $c\in\mathbb{N}$ and $i\in\{r,b\}$]. Finally, symmetry with Eq.~\eqref{eq:hrn_Rcbrc} implies that
\begin{align}
  R(c,b|c,r) & = \frac{w_c(\langle k \rangle_c - c )/\langle k \rangle - \sum_{c^{\prime\prime}<c} C_{cc^{\prime\prime}}}{w_cc/\langle k \rangle} \ ,
\end{align}
and normalization leads to
\begin{align}
  R(c,r|c,r) & = \frac{2w_cc/\langle k \rangle - C_{cc} - 2\sum_{c^{\prime\prime}>c}C_{cc^{\prime\prime}}}{w_cc/\langle k \rangle} \ ,
\end{align}
\end{subequations}
where we have used the fact that $\sum_{c^{\prime\prime}} C_{cc^{\prime\prime}} = w_c\langle k \rangle_c/\langle k \rangle$.

To compute the size of the giant component in the limit of large networks ($N \rightarrow \infty$), we define a probability generating function (pgf)
\begin{align}
  g_c(\bm{x}) & = \sum_{\bm{k}} P_c(\bm{k}) \prod_{i} \Big[ (1-T) + T \sum_{c^\prime,j}R(c^{\prime},j|c,i) x_{c^{\prime}j}\Big]^{k_i}
\end{align}
that generates the distribution of the number of nodes of each type (i.e., coreness $c^{\prime}$) that can be reached from a node of coreness $c$ (the subscript $j$ of the variable $x_{c^\prime j}$ indicates the color of the stubs from which the node has been reached). Similarly, let us consider a node of coreness $c$ that has been reached from one of its stubs, the distribution of the number and type of its other neighbors (its \textit{excess} degree distribution) is generated by one of the two following pgfs
\begin{align}
  f_{cr}(\bm{x}) & = \sum_{\bm{k}} P_c(\bm{k}) \prod_{i} \Big[ 1-T + T \sum_{c^\prime,j}R(c^{\prime},j|c,i) x_{c^{\prime}j}\Big]^{k_i-\delta_{ir}} \\
  f_{cb}(\bm{x}) & = \sum_{\bm{k}} \frac{k_b P_c(\bm{k})}{\langle k \rangle_c - c} \prod_{i} \Big[ 1-T + T \sum_{c^\prime,j}R(c^{\prime},j|c,i) x_{c^{\prime}j}\Big]^{k_i-\delta_{ib}}
\end{align}
depending on the color of the stubs from which the node has been reached. The size of the giant component is then given by (see Ref.~\cite{Allard13b} for a complete and more general theoretical framework)
\begin{align} \label{eq:hrn_pgf_order_param}
  S = 1 - \sum_{c} w_c g_c(\bm{a})
\end{align}
where $\bm{a}\equiv\{a_{ci}\}_{c\in\mathbb{N},i\in\{r,b\}}$ is the probability that a node of coreness $c$ reached by one of its stubs of color $i$ does not belong to the giant component. These probabilities correspond to the stable fixed point of the system of equations
\begin{align} \label{eq:hrn_pgf_fixed_point}
  a_{ci} & = f_{ci}(\bm{a})
\end{align}
with $c\in\mathbb{N}$ and $i\in\{r,b\}$. As the distributions generated by $f_{ci}(\bm{x})$ are normalized, $\bm{a}=\bm{1}$ is always a solution of Eqs~\eqref{eq:hrn_pgf_fixed_point} and corresponds to the subcritical regime $S=0$. At $T=T_\mathrm{c}$, this fixed point undergoes a transcritical bifurcation and looses its stability to another solution in $[0,1)^{c_\mathrm{max}}$. This supercritical regime corresponds to the existence of a giant component ($S>0$); the critical point $T_\mathrm{c}$ is obtained from a stability analysis of Eqs.~\eqref{eq:hrn_pgf_fixed_point} around $\bm{a}=\bm{1}$.
\begin{figure}[t]
  \includegraphics[trim = 0mm 4mm 0mm 0mm, clip, height=3.95cm]{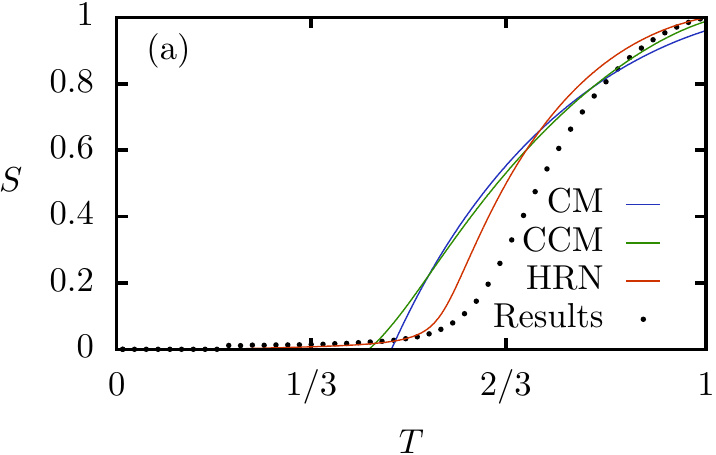} \\
  \includegraphics[trim = 0mm 0mm 0mm 0mm, clip, height=4.345cm]{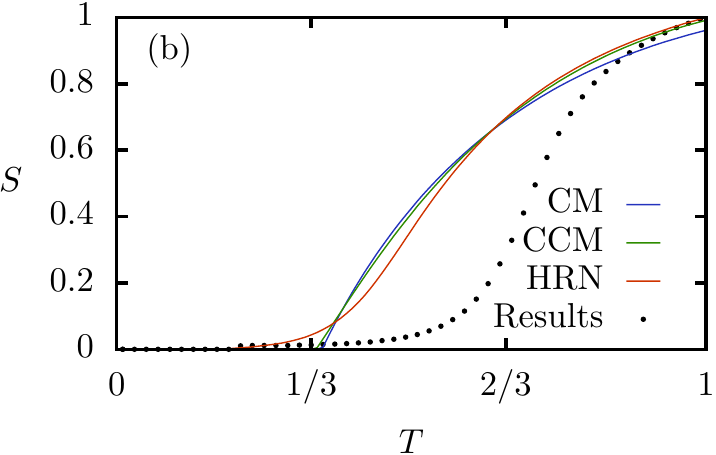} \\
  \caption{\label{fig:hrn_results_2}Results of bond percolation on real networks (black) compared with analytical predictions obtained with the CM (blue), CCM (green) and HRN (red). The networks are: (a) a subset of the power grid of Poland, and (b) the Western States Power Grid of the United States. See Table~\ref{tab:hrn_networks} for further details.}
\end{figure}
\subsection{Numerical HRN networks} \label{sec:hrn_hrn_algo}
To generate networks with a given core structure, we start with $N \gg 1$ nodes whose number of stubs is drawn from the degree distribution $\{P(k)\}_{k\in\mathbb{N}} = \{\sum_{c} K_{ck}\}_{k\in\mathbb{N}}$, and randomly match stubs to create edges (as done for the CM \cite{Newman02_PhysRevE}). Next, for each node, we assign a coreness $c$ with probability $Q_k(c) = K_{ck}/P(k)$; $c$ of its stubs are then randomly selected as red and the $k-c$ others are identified as blue. Finally, we apply the following Metropolis-Hastings rewiring algorithm (similar to the one proposed in Ref.~\cite{Newman02_PhysRevLett}). At each step, two edges are randomly selected: edge 1 joins nodes of coreness $c_1$ and $c_1^\prime$ via their respective stubs of color $i_1$ and $j_1$ ($c_2$, $i_2$, $c_2^\prime$ and $j_2$ for edge 2). We replace these two edges by edge 3 ($c_1$, $i_1$, $c_2$ and $i_2$) and edge 4 ($c_1^\prime$, $j_1$, $c_2^\prime$ and $j_2$) with probability
\begin{align*}
  \min\left\{1,\frac{\Gamma(c_1,i_1;c_2,i_2)\Gamma(c_1^\prime,j_1;c_2^\prime,j_2)}{\Gamma(c_1,i_1;c_1^\prime,j_1)\Gamma(c_2,i_2;c_2^\prime,j_2)}\right\} \ ,
\end{align*}
where $\Gamma(c,i;c^\prime,j)$ is the wanted fraction of edges that join nodes of coreness $c$ and $c^\prime$ via their respective stubs of color $i$ and $j$. These fractions are readily obtained from the matrix $\mathbf{C}$ [joint probabilities of Eqs.~\eqref{eq:hrn_R}]
\begin{equation}
  \begin{aligned}
    \Gamma(c,r;c^\prime,b) = \Gamma(c^\prime,b;c,r) & = C_{cc^\prime} \\
    \Gamma(c,r;c,b) = \Gamma(c,b;c,r) & = w_c(\langle k \rangle_c - c )/\langle k \rangle - \sum_{c^{\prime\prime}<c} C_{cc^{\prime\prime}} \\
    \Gamma(c,r;c,r) & = 2w_cc/\langle k \rangle - C_{cc} - 2\sum_{c^{\prime\prime}>c} C_{cc^{\prime\prime}}
  \end{aligned}
\end{equation}
where $c<c^\prime$, and $\Gamma(c,i;c^\prime,j)$ is zero for all other combinations. This procedure preserves the degree distribution, and up to finite-size constraints, has the wanted core structure as its fixed point and is ergodic over the ensemble of networks defined by the HRN model. Figure~\ref{fig:hrn_validation} compares the predictions of Eqs.~\eqref{eq:hrn_pgf_order_param}--\eqref{eq:hrn_pgf_fixed_point} with the size of the giant component found in networks generated through this algorithm and shows a perfect agreement.
\begin{figure}[t]
  \includegraphics[width=\linewidth]{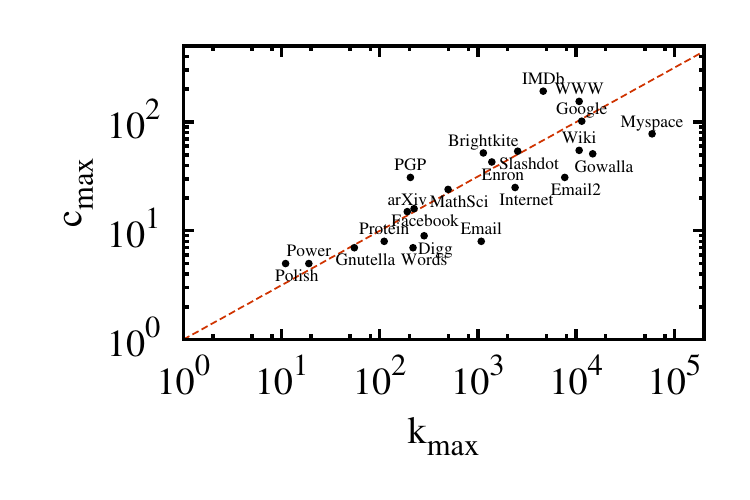}
  \caption{\label{fig:hrn_cbyk}Relation between the highest coreness, $c_\mathrm{max}$, and the highest degree, $k_\mathrm{max}$, for different real networks. The dashed line corresponds to $c_\mathrm{max} \propto \sqrt{k_\mathrm{max}}$.}
\end{figure}
\begin{table*}
  \centering
  \caption{\label{tab:hrn_networks}Description and properties of the real networks used in Figs.~\ref{fig:hrn_validation}--\ref{fig:hrn_cbyk}.}
  \begin{tabular}{l | c | c | c | c | c | c}
    \hline
    \hline
    \multicolumn{1}{c|}{Description} & \multicolumn{1}{c|}{$N$} & \multicolumn{1}{c|}{$\langle k \rangle$} & \multicolumn{1}{c|}{$k_\mathrm{max}$} & \multicolumn{1}{c|}{$c_\mathrm{max}$} & Fig. & Ref.  \\
    \hline
    Web of trust of the Pretty Good Privacy (PGP) encryption algorithm       &  10 680 & 4.55 &    205 & 31 & \ref{fig:hrn_results_1}(c), \ref{fig:hrn_cbyk} & \cite{Boguna04_PhysRevE}           \\
    Snapshot of the Gnutella peer-to-peer network                            &  36 682  & 4.82  & 55 & 7 & \ref{fig:hrn_results_1}(a), \ref{fig:hrn_cbyk}  & \cite{Ripeanu02_PeerToPeerSystems} \\
    Large subset of the Facebook social network                              & 63 891  & 5.74  &  223   & 16 & \ref{fig:hrn_results_1}(f), \ref{fig:hrn_cbyk} & \cite{Viswanath09_WOSN}            \\
    Snapshot of the Gowalla location-based social network                    & 196 591  & 9.67  & 14 730 & 51 & \ref{fig:hrn_results_1}(b), \ref{fig:hrn_cbyk} & \cite{Cho11_KDD}                   \\
    Email exchange network from an undisclosed European institution &  300 069  & 2.80  & 7 631 & 31 & \ref{fig:hrn_validation}, \ref{fig:hrn_cbyk} & \cite{Leskovec07_TKDD}             \\
    Subset of the World Wide Web                                             & 325 729 & 6.69 & 10 721 & 155 & \ref{fig:hrn_results_1}(d), \ref{fig:hrn_cbyk} & \cite{Barabasi99_Science}          \\
    Co-authorship network of MathSciNet before 2008                          &  391 529  &  4.46 & 496 & 24 & \ref{fig:hrn_validation}, \ref{fig:hrn_results_1}(e), \ref{fig:hrn_cbyk} & \cite{Palla08_NewJPhys}            \\
    &&&&&&\\
    Subset of the power grid of Poland                                       &   3 374 & 2.41 & 11     & 5 & \ref{fig:hrn_validation}, \ref{fig:hrn_results_2}(a), \ref{fig:hrn_cbyk} & \cite{Zimmerman2011}               \\
    Western States Power Grid of the United States                           &   4 941 & 2.67 &     19 & 5 & \ref{fig:hrn_results_2}(b), \ref{fig:hrn_cbyk}  & \cite{Watts98_Nature}              \\
    &&&&&&\\   
    Email communication within the University Rovira i Virgili & 1 134 & 9.07 & 1 080 & 8 & \ref{fig:hrn_cbyk} & \cite{Palla05_Nature} \\
    Protein-protein interactions in \textit{S. cerevisiae}					& 2 640 & 5.00 & 111 & 8 & \ref{fig:hrn_cbyk} & \cite{Palla05_Nature} \\
    Word association graph from the South Florida Free Association norms & 7 207 & 8.82 & 218 & 7 & \ref{fig:hrn_cbyk} & \cite{Palla05_Nature} \\
    Network of hyperlinks between Google's webpages & 15 763 & 18.96 & 11 401 & 102 & \ref{fig:hrn_cbyk} & \cite{Farkas_directednetwork} \\
    Structure of the Internet at the level of autonomous systems & 22 963 & 4.22 & 2390 & 25 & \ref{fig:hrn_cbyk} & \cite{Hebert-Dufresne11_PhysRevLett} \\
    Reply network of the social news website Digg               & 30 398 & 5.60 & 283 & 9 & \ref{fig:hrn_cbyk} & \cite{b565} \\
    The cond-mat arXiv co-authorship network circa 2005   					& 30 561 & 8.24 & 191 & 15 & \ref{fig:hrn_cbyk} & \cite{Palla05_Nature} \\
    Email interchanges between different Enron email addresses & 36 692 & 10.02 & 1 383 & 43 & \ref{fig:hrn_cbyk} & \cite{Klimt2004} \\
    Brightkite location-based online social network & 58 228 & 7.35 & 1 134 & 52 & \ref{fig:hrn_cbyk} & \cite{Cho11_KDD} \\
    Network of tagged relationships on the Slashdot news website & 77 360 & 12.13 & 2 539 & 54 & \ref{fig:hrn_cbyk} & \cite{Leskovec_08} \\
    Friendships between 100 000 Myspace accounts & 100 000 & 16.82 & 59 108 & 78 & \ref{fig:hrn_cbyk} & \cite{Ahn:2007:ATC:1242572.1242685} \\
    Network of interactions between the users of the English Wikipedia & 138 592 & 10.33 & 10 715 & 55 & \ref{fig:hrn_cbyk} & \cite{konect:maniu2011}\\
    Co-acting network in movies released after December 31st 1999 & 716 463 & 21.40 & 4625 & 192 & \ref{fig:hrn_cbyk} & \cite{Hebert-Dufresne11_PhysRevLett} \\
    \hline 
    \hline
  \end{tabular}
\end{table*}
\subsection{Results}
Figures~\ref{fig:hrn_results_1}--\ref{fig:hrn_results_2} display the predictions of Eqs.~\eqref{eq:hrn_pgf_order_param}--\eqref{eq:hrn_pgf_fixed_point} with the size of the giant component found in real networks (see caption and Table~\ref{tab:hrn_networks} for a complete description), and with the predictions of the CM and the CCM. These particular networks were chosen to highlight some important results. 

First, we find that the HRN model performs \textit{at least as well} as the CCM in all investigated cases. This observation is interesting as the HRN model requires less input information than the CCM. Indeed the required information scales roughly as $k_\mathrm{max} c_\mathrm{max} + c_\mathrm{max}^2$. As shown in Fig.~\ref{fig:hrn_cbyk}, $c_\mathrm{max}$ scales approximately as $k_\mathrm{max}^{1/2}$ in many real networks, hence the input information in the HRN model scales roughly as $k_\mathrm{max}^{3/2}$. Considering the fact that $k_\mathrm{max}$ in real networks is often well above 10\textsuperscript{2} (see Table~\ref{tab:hrn_networks}), this difference results in a much faster computation and a major memory gain. Moreover, this implies that, although the HRN model does not account explicitly for the degree-degree correlations, they are effectively captured by the matrices $\mathbf{K}$ (degree-coreness correlations) and $\mathbf{C}$ (coreness-coreness correlations). As shown on Figs.~\ref{fig:hrn_results_1}--\ref{fig:hrn_results_2}, this effect was observed on all available real-world networks.

Second, and perhaps surprisingly, we see in Fig.~\ref{fig:hrn_results_2}(a) that the ``S'' shape obtained from the Polish power grid, typically due to finite size, is well reproduced by the HRN model, which is formally infinite in size. More precisely, this shape is usually attributed to the finite size of the network ($N=3374$ for the Polish power grid) as the small components---whose average size formally diverges at $T=T_c$---are misinterpreted as an emerging giant component. Interestingly, the results from the HRN model suggest that this shape is not a numerical artifact of the percolation algorithm, but that it is rather a signature of its geographically-embedded nature due to strong \textit{coreness-related} correlations. This unexpected property of the HRN model is confirmed on another, more clustered, power grid on Fig.~\ref{fig:hrn_results_2}(b). In this case, adding clustering to the HRN is expected to shift its prediction towards higher values of $T$, i.e., closer to the results from the real network. In fact, the HRN model is more accurate in predicting percolation on the Polish power grid (clustering coefficient $C=0.02$) than for this grid ($C=0.08$). A clustered version of the HRN model seems to offer a promising avenue for the modeling of geographically-embedded networks such as power grids.

In this regard, the results of Figs.~\ref{fig:hrn_results_1}(e)--(f) add even more emphasis on the importance of including the effect of clustering in a subsequent version of the HRN model. Indeed, co-authorship networks \ref{fig:hrn_results_1}(e) are notoriously clustered networks as authors of a same paper are all connected via a fully-connected clique. Similarly, in Facebook \ref{fig:hrn_results_1}(f), people belonging to a same social group (e.g., classmates, colleagues, teammates) tend all to be connected to one another, yielding almost fully-connected cliques. Again, we expect in this situation that clustering would reduce the size of the giant component (due to redundant connections in cliques), hence bringing the predictions of a clustered HRN model closer to the behaviors observed with the real networks.
\section{Conclusion}
We have shown that the core structure can be useful beyond the characterization and visualization of networks. It serves well modeling efforts and is efficient in reproducing the structural properties of real networks. Moreover, a few simple connection rules can enforce a core structure in random networks for which the outcome of bond percolation can be predicted with the well-established pgf approach\footnote{Codes solving the theoretical model and generating the networks are available at \texttt{http://dynamica.phy.ulaval.ca}.}. We feel that this work sets the stage for further improvements (specifically the inclusion of clustering) and paves the way towards a more complete analytical description of percolation on real networks.
%
%
%
%
%
%
\begin{acknowledgments}
 The authors would like to acknowledge the financial support of the Canadian Institutes of Health Research, the Natural Sciences and Engineering Research Council of Canada, and the Fonds de recherche du Qu\'ebec--Nature et technologies.
\end{acknowledgments}
%
%
\appendix
\section{Configuration Model\label{app:CM}}
The most influential quantity with regard to bond percolation on networks is the degree distribution: the distribution of the number of connections (degree) that nodes have. 
The simplest analytical model that incorporates an arbitrary degree distribution is the CM \cite{Newman01_PhysRevE}. It defines a maximally random network ensemble that is random in all respects other than the degree distribution $\{P(k)\}_{k\in\mathbb{N}}$: the probability for a randomly chosen node to have a degree equal to $k$. Networks of this ensemble are generated by creating a set of $N$ nodes, each with a number of stubs drawn from the degree distribution, and then by pairing randomly stubs to form edges. \\

To compute the size $S^{\mathrm{CM}}$ of the giant component and the value $T_c^{\mathrm{CM}}$ of the percolation threshold, we define the probability generating function \cite{Newman02_PhysRevE}
\begin{align}
 g(x) = \sum_{k=0}^{\infty} P(k) [(1-T) + Tx]^k
\end{align}
that generates the degree distribution. The first derivative of $g(x)$ evaluated at $x=1$ corresponds to the average degree of the nodes $g^\prime(1) = \langle k \rangle$. We also define
\begin{align}
 f(x) = \frac{g^\prime(x)}{g^\prime(1)} = \frac{1}{\langle k \rangle} \sum_{k^\prime=1}^{\infty} k^\prime P(k^\prime) [(1-T) + Tx]^{k^\prime-1}
\end{align}
that generates the number of \textit{other} neighbors of a node that has been reached by following a randomly chosen edge (i.e., the \textit{excess} degree distribution). The size of the giant component is directly obtained via
\begin{align}
 S^{\mathrm{CM}} & = 1 - g(a^{\mathrm{CM}}) \ ,
\end{align}
where $a^{\mathrm{CM}}$ is the probability that a randomly chosen edge does not lead to the giant component. It is the stable fixed point of
\begin{align}
 a^{\mathrm{CM}} & = f(a^{\mathrm{CM}})
\end{align}
in $[0,1]$. The solution $a^{\mathrm{CM}}=1$ corresponds to the absence of a giant component ($S^{\mathrm{CM}}=0$). The percolation threshold is the point at which this solution becomes unstable.

To model bond percolation on a given network with the CM, one simply has to extract the degree distribution; the required information therefore scales as $k_{\mathrm{max}}$, the highest degree of the network. The original network is then found within the network ensemble generated by the CM, the ensemble composed of all possible networks one could design with the exact same degree distribution.

\section{Correlated Configuration Model\label{app:CCM}}
Apart from the degree distribution, real networks are typically characterized by strong correlations regarding \textit{who is connected with whom}.

One way to include such correlations into a random network model is through the \textit{joint degree distribution} $\{P(k,k^\prime)\}_{k,k^\prime\in\mathbb{N}}$ giving the probability that a randomly chosen edge has nodes of degree $k$ and $k^\prime$ at its ends. This yields a \textit{Correlated Configuration Model} (CCM) that defines a maximally random network ensemble having arbitrary degree-degree correlations with a corresponding degree distribution. The degree distribution is encoded in $\{P(k,k^\prime)\}_{k,k^\prime\in\mathbb{N}}$ through the identity
\begin{align}
  \sum_{k^\prime} P(k,k^{\prime}) = \frac{kP(k)}{\langle k \rangle} \ .
\end{align}

Generating networks from this ensemble proceeds as for the CM: $N$ nodes, whose degrees are drawn from $\{P(k)\}_{k\in\mathbb{N}}$, are connected via the stub pairing scheme. A Metropolis-Hastings rewiring algorithm \cite{Newman02_PhysRevLett} is then applied whose fixed point is the network ensemble defined by $\{P(k,k^\prime)\}_{k,k^\prime\in\mathbb{N}}$. At each step, two edges are randomly chosen: edge 1 joins nodes $m_1$ and $n_1$ with respective degree $i_1$ and $j_1$ ($m_2$, $n_2$, $i_2$ and $j_2$ for edge 2). These two edges are replaced by edge 3 ($m_1$, $m_2$, $i_1$ and $i_2$) and edge 4 ($n_1$, $n_2$, $j_1$ and $j_2$) with probability
\begin{align}
  \min\left\{ 1, \frac{P(i_1,i_2) P(j_1,j_2)}{P(i_1,j_1) P(i_2,j_2)} \right\} \ .
\end{align}

The size $S^{\mathrm{CCM}}$ of the giant component is computed as in the CM \cite{Newman02_PhysRevLett}
\begin{align}
  S^{\mathrm{CCM}} = 1 - \sum_{k=0}^{\infty} P(k) [(1-T) + Ta_{k}]^{k} = 1 - g(\bm{a})
\end{align}
where $\bm{a} = \{a_k\}_{k\in\mathbb{N}}$ is the set of probabilities that an edge leading toward a node with a degree $k$ is not attached to the giant component. They correspond to the stable fixed point in $[0,1]^{k_\mathrm{max}}$ of the system of equations
\begin{align} \label{eq:ccm_fixed_point}
  a_k = \frac{\sum_{k^\prime} P(k,k^\prime) [(1-T) + Ta_{k^\prime}]^{k^\prime-1}}{\sum_{k^\prime} P(k,k^\prime)}
\end{align}
with $k\in\mathbb{N}$. The value $T_c$ of the percolation threshold is the value for which the fixed point $\bm{a}=\bm{1}$ of Eqs.~\eqref{eq:ccm_fixed_point} becomes unstable. \\

To model bond percolation on a given network with the CCM, one simply has to extract the joint degree distribution. This is achieved by scanning the degree of the two nodes at the end of each edge of the network; the required information therefore scales as $k_{\mathrm{max}}^2$. The original network is then found within the random network ensemble of all networks with the same degree distribution and degree-degree correlations. Note that this ensemble is a subset of the ensemble generated by the CM with the same degree distribution.

\section{Consistency conditions on $\mathbf{K}$ and $\mathbf{C}$\label{app:cons}}
The consistency conditions on the matrices $\mathbf{K}$ and $\mathbf{C}$ can be summarized as follows: they must encode an ensemble of \textit{closed} networks. In other words, \textit{all stubs must be paired}, and this must be done in accordance with the stubs matching rules (e.g., two blue stubs cannot be paired). Consequently, there is no k-core structure that the HRN model cannot model as long as it is realistic. This will always be the case when $\mathbf{K}$ and $\mathbf{C}$ are extracted from real networks. \\

First, there must be as many edges leaving nodes of coreness $c$ toward nodes of coreness $c^\prime$ as there are in the opposite direction. This requires that $C_{c c^\prime} = C_{c^\prime c}$, a condition that is always fulfilled since $\mathbf{C}$ is defined as a symmetric matrix
\begin{align}
  \mathbf{C} = \mathbf{C}^\mathrm{T} \ .
\end{align}
Secondly, the degree of each node is bounded from below by its coreness, hence
\begin{align}
  K_{ck} = 0 \text{\qquad for } k<c \ .
\end{align}
Thirdly, both $\mathbf{K}$ and $\mathbf{C}$ must prescribe the same number of stubs stemming from nodes of coreness $c$
\begin{align}
 \sum_{k} kK_{ck} = \langle k \rangle \sum_{c^\prime} C_{cc^\prime} \ ,
\end{align}
where the extra factor $\langle k \rangle$ accounts for the fact that $\mathbf{K}$ ``counts'' nodes whereas $\mathbf{C}$ ``counts'' stubs (i.e., multiplying both sides by the number of nodes $N$ yields absolute numbers instead of \textit{per capita} averages). Finally, as the coreness of the nodes defines their number of red stubs, the matrix $\mathbf{C}$ is subjected to the following additional constraints for every $c$
\begin{align} \label{eq:hrn_sm_condition4}
  \langle k \rangle \sum_{c^\prime>c} C_{cc^\prime} \leq w_cc \leq \langle k \rangle \sum_{c^\prime\geq c} C_{cc^\prime} \ .
\end{align}
The first inequality states that there must be at least as many red stubs stemming from nodes of coreness $c$ as there are edges leaving the $c$-shell toward nodes of higher coreness. Equality then means that all red stubs lead to node of higher coreness. The second inequality states that all red stubs must lead to nodes of coreness $c$ or higher. Equality occurs when all blue stubs are directed toward nodes of coreness $c^\prime<c$. A similar expression to \eqref{eq:hrn_sm_condition4} can be derived for blue stubs
\begin{align}
  \langle k \rangle \sum_{c^\prime<c} C_{cc^\prime} \leq (\langle k \rangle_c - c) w_c \leq \langle k \rangle \sum_{c^\prime\leq c} C_{cc^\prime} \ ,
\end{align}
and can be interpreted analogously.

%
%
%
%
%
%
%
%
%
\end{document}